\documentclass[aps,prl,reprint,superscriptaddress,floatfix]{revtex4-2}

\usepackage{amsmath,amssymb}
\usepackage{graphicx}
\usepackage{bm}
\usepackage{microtype}
\usepackage[hidelinks]{hyperref}

\begin{document}

\title{Experimental Evidence for MeV Undulator Radiation Generated by Ultrarelativistic Positrons in a Periodically Bent Crystal}

\author{L. Bandiera}
\email{bandiera@fe.infn.it}
\affiliation{INFN Sezione di Ferrara, Ferrara, Italy}

\author{N. Canale}
\affiliation{INFN Sezione di Ferrara, Ferrara, Italy}
\affiliation{Dipartimento di Fisica e Scienze della Terra, Universit\`a degli Studi di Ferrara, Ferrara, Italy}

\author{S. Carsi}
\affiliation{Nuclear Instruments S.r.l., Lambrugo, Como, Italy}

\author{M. Cavallina}
\affiliation{INFN Sezione di Ferrara, Ferrara, Italy}

\author{F. Cescato}
\affiliation{INFN Sezione di Ferrara, Ferrara, Italy}
\affiliation{Dipartimento di Fisica e Scienze della Terra, Universit\`a degli Studi di Ferrara, Ferrara, Italy}

\author{N. Charitonidis}
\affiliation{CERN, Geneva, Switzerland}

\author{D. De Salvador}
\affiliation{Universit\`a degli Studi di Padova, Padua, Italy}
\affiliation{INFN Laboratori Nazionali di Legnaro, Legnaro, Italy}

\author{M. Diaz Castro}
\affiliation{Higher Institute of Technologies and Applied Sciences, University of Havana (InSTEC-UH), Havana, Cuba}

\author{E. Diociaiuti}
\affiliation{INFN Laboratori Nazionali di Frascati, Frascati, Italy}

\author{P. Fedeli}
\affiliation{INFN Sezione di Ferrara, Ferrara, Italy}
\affiliation{Dipartimento di Fisica e Scienze della Terra, Universit\`a degli Studi di Ferrara, Ferrara, Italy}

\author{L. G. Foggetta}
\affiliation{LASA - INFN Milano, Milano, Italy}

\author{A. Forlani}
\affiliation{INFN Sezione di Ferrara, Ferrara, Italy}
\affiliation{Dipartimento di Fisica e Scienze della Terra, Universit\`a degli Studi di Ferrara, Ferrara, Italy}

\author{A. Gianoli}
\affiliation{INFN Sezione di Ferrara, Ferrara, Italy}

\author{V. Guidi}
\affiliation{INFN Sezione di Ferrara, Ferrara, Italy}
\affiliation{Dipartimento di Fisica e Scienze della Terra, Universit\`a degli Studi di Ferrara, Ferrara, Italy}

\author{L. Malagutti}
\affiliation{INFN Sezione di Ferrara, Ferrara, Italy}

\author{F. Mancarella}
\affiliation{CNR - ISMN, Bologna, Italy}

\author{S. Mangiacavalli}
\affiliation{Universit\`a degli Studi dell'Insubria, Como, Italy}
\affiliation{INFN Sezione di Milano Bicocca, Milan, Italy}

\author{A. Mazzolari}
\affiliation{INFN Sezione di Ferrara, Ferrara, Italy}
\affiliation{Dipartimento di Fisica e Scienze della Terra, Universit\`a degli Studi di Ferrara, Ferrara, Italy}

\author{R. Negrello}
\affiliation{INFN Sezione di Ferrara, Ferrara, Italy}
\affiliation{Dipartimento di Fisica e Scienze della Terra, Universit\`a degli Studi di Ferrara, Ferrara, Italy}

\author{G. Patern\`o}
\affiliation{INFN Sezione di Ferrara, Ferrara, Italy}

\author{M. Prest}
\affiliation{Universit\`a degli Studi dell'Insubria, Como, Italy}
\affiliation{INFN Sezione di Milano Bicocca, Milan, Italy}

\author{J. Reyes Garrido}
\affiliation{Sapienza University of Rome, Rome, Italy}
\affiliation{INFN Sezione di Ferrara, Ferrara, Italy}

\author{M. Romagnoni}
\affiliation{INFN Sezione di Ferrara, Ferrara, Italy}
\affiliation{Dipartimento di Fisica e Scienze della Terra, Universit\`a degli Studi di Ferrara, Ferrara, Italy}

\author{A. Saputi}
\affiliation{INFN Sezione di Ferrara, Ferrara, Italy}

\author{F. Sgarbossa}
\affiliation{Universit\`a degli Studi di Padova, Padua, Italy}
\affiliation{INFN Laboratori Nazionali di Legnaro, Legnaro, Italy}

\author{A. Sytov}
\affiliation{INFN Sezione di Ferrara, Ferrara, Italy}

\author{V. V. Tikhomirov}
\noaffiliation

\author{E. Vallazza}
\affiliation{INFN Sezione di Milano Bicocca, Milan, Italy}

\author{D. Valzani}
\affiliation{Universit\`a degli Studi di Padova, Padua, Italy}
\affiliation{INFN Laboratori Nazionali di Legnaro, Legnaro, Italy}

\begin{abstract}
We report evidence for undulator radiation generated by ultrarelativistic positrons 
in a large-amplitude, long-period crystalline undulator. A pattern of Si$_3$N$_4$ 
stressors periodically bend the (110) planes of a Silicon crystal with a geometry 
measured independently by x-ray diffraction. Under planar alignment, the forward 
energy loss showed a localized excess at $6.8 \pm 0.3$~MeV, matching 
the predictions of simulations. Experimental results also indicate that an intense 
positron beam could make the undulator competitive in brilliance with 
state-of-the-art inverse-Compton gamma-ray sources.
\end{abstract}

\maketitle

The generation of intense, tunable photon beams at sub-\AA{} wavelengths remains a major goal in modern accelerator and radiation physics. High-brilliance radiation in the hard-X-ray and gamma-ray domains is required for applications ranging from nuclear photonics and nuclear-structure studies to particle physics and advanced imaging. Conventional magnetic undulators, whose periods are typically on the centimeter scale, efficiently generate radiation up to the hard-X-ray domain. Extending this concept toward much higher photon energies, however, requires increasingly energetic electron beams and large accelerator and undulator systems. Laser-assisted schemes, such as inverse Compton scattering~\cite{Krafft2010,Petrillo2023} and laser--ion concepts including the Gamma Factory~\cite{Placzek2019,Budker2022}, offer alternative routes to gamma-ray production, but generally rely on high-performance beams combined with intense, high-repetition-rate laser systems.

\begin{figure}[t]
\centering
\includegraphics[width=\columnwidth]{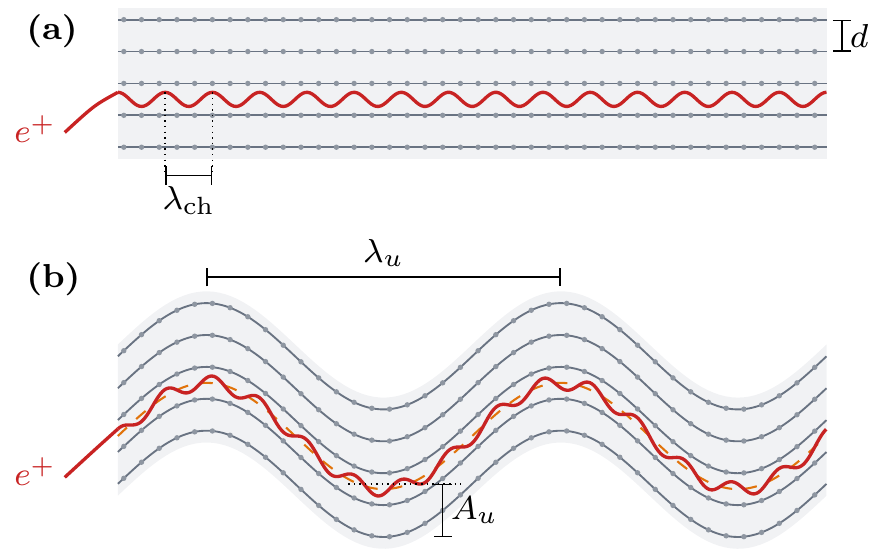}
\caption{(a) Scheme of positron channeling in a straight crystal. (b) Scheme of positron channeling in a periodically bent crystal to realize a CU.}
\label{undulator}
\end{figure}

Crystalline undulators (CU) offer a distinct approach. In an aligned crystal, the interplanar potential can confine charged particles~\cite{Lindhard1965}, giving rise to microscopic channeling oscillations and their associated radiation~\cite{Kumakhov1976} (see Fig. \ref{undulator}a). In the original CU concept, crystallographic planes are additionally bent with a controlled spatial periodicity (see Fig. \ref{undulator}b). A channeled particle then follows the periodically curved planes, so that the bending motion is superimposed on the intrinsic channeling oscillations and generates an additional radiation component~\cite{Kaplin1980,Baryshevsky1980}. Whereas the characteristic scale of channeling radiation is determined primarily by the microscopic interplanar potential, CU radiation introduces additional geometrical parameters set by the engineered deformation, namely the bending period $\lambda_u$ and amplitude $A_u$. Since crystalline bending periods can be orders of magnitude shorter than those of magnetic undulators, this additional radiation component can reach the MeV range at multi-GeV beam energies.

This concept was developed into the large-amplitude, long-period (LALP) regime, characterized by $A_u\gg d$ and $\lambda_u\gg\lambda_{\rm ch}$, where $d$ is the interplanar spacing and $\lambda_{\rm ch}$ is the characteristic channeling-oscillation period~\cite{Korol1999,Korol2014,Korol-2020,sushko-2022}. In this regime, the particle follows the periodically curved planes while undergoing much smaller intrinsic channeling oscillations. The engineered bending therefore introduces a distinct transverse motion and a corresponding radiation component on a spectral scale separated from that of channeling radiation. Under suitable conditions, the CU contribution can become comparable to or exceed the intrinsic channeling-radiation component.

Realizing the large-amplitude regime experimentally is demanding because a sufficiently large and uniform periodic bulk deformation must be produced while preserving stable channeling throughout the active volume. Different technologies have been explored, including acoustically bent
crystals~\cite{kaleris-2024}, graded Si$_{1-x}$Ge$_x$
structures~\cite{Mikkelsen2000,Backe2013}, surface grooving~\cite{Bellucci2003,Bagli2014,Camattari2019},
and periodically patterned surface stresses~\cite{Avakian2002,Guidi2007,Malagutti2025}. 
An early measurement with $10$~GeV positrons at U70-IHEP in a grooved Si crystal
revealed a spectral enhancement qualitatively consistent with an undulator
contribution~\cite{Baranov2005}. However, the limited signal-to-background ratio and
the possible overlap with channeling radiation prevented an unambiguous identification
of the undulator component.

Complementary approaches have explored small-amplitude periodic perturbations,
$A_u \ll d$, first through resonant coupling between the induced motion and
channeling radiation~\cite{Ikezi1984,Dedkov1994}. This line of research later
led to the small-amplitude, short-period (SASP) regime proposed in
Ref.~\cite{Kostyuk2013,korol-2023} and investigated experimentally with strained
Si$_{1-x}$Ge$_x$ crystals using electron beams at MAMI~B~\cite{Wistisen2014}.
More recently at MAMI, a four-period boron-doped diamond superlattice produced a
narrow radiation feature around $1.3$~MeV with $855$~MeV electrons~\cite{Backe2026}.
This device featured a sub-interplanar modulation amplitude with a $5.0~\mu$m
period, intermediate between the SASP and LALP regimes.
Thus, these experiments have explored electron-based configurations with small-amplitude modulations, rather than the conventional LALP regime. Positrons offer a particularly favorable route to CU. Indeed, under
channeling, positrons remain confined between atomic planes, i.e. they
experience weaker nuclear scattering than for electrons and therefore exhibit much longer
dechanneling lengths. This enables the use of thicker crystals with many more undulator
periods, resulting in higher intensity and narrower spectral lines, which both contribute to
the generation of high-brilliance photon beams. Although high-energy positron beams are
currently available at fewer facilities than for electron beams, upcoming sources such as
the FCC-ee injector \cite{Agapov2026} and the proposed CEBAF positron upgrade at Jefferson Lab \cite{Accardi:2020eplusJLab} will provide intense, high-quality positron beams, opening a clear path
toward positron-driven CUs as compact, high-brilliance MeV photon
sources.

Here we report evidence for CU radiation in the LALP regime using a ten-period, stressor-engineered Si(110) CU and a $35$~GeV positron beam. The periodic bulk deformation is characterized independently of the radiation measurement by combining high-resolution X-ray diffraction with finite-element modeling. The resulting geometry fixes the expected radiation energy before comparison with the photon data, enabling a direct quantitative test of the connection between crystal geometry, relativistic particle dynamics, and MeV photon emission.

\begin{figure*}[t]
\centering
\includegraphics[width=0.94\textwidth]{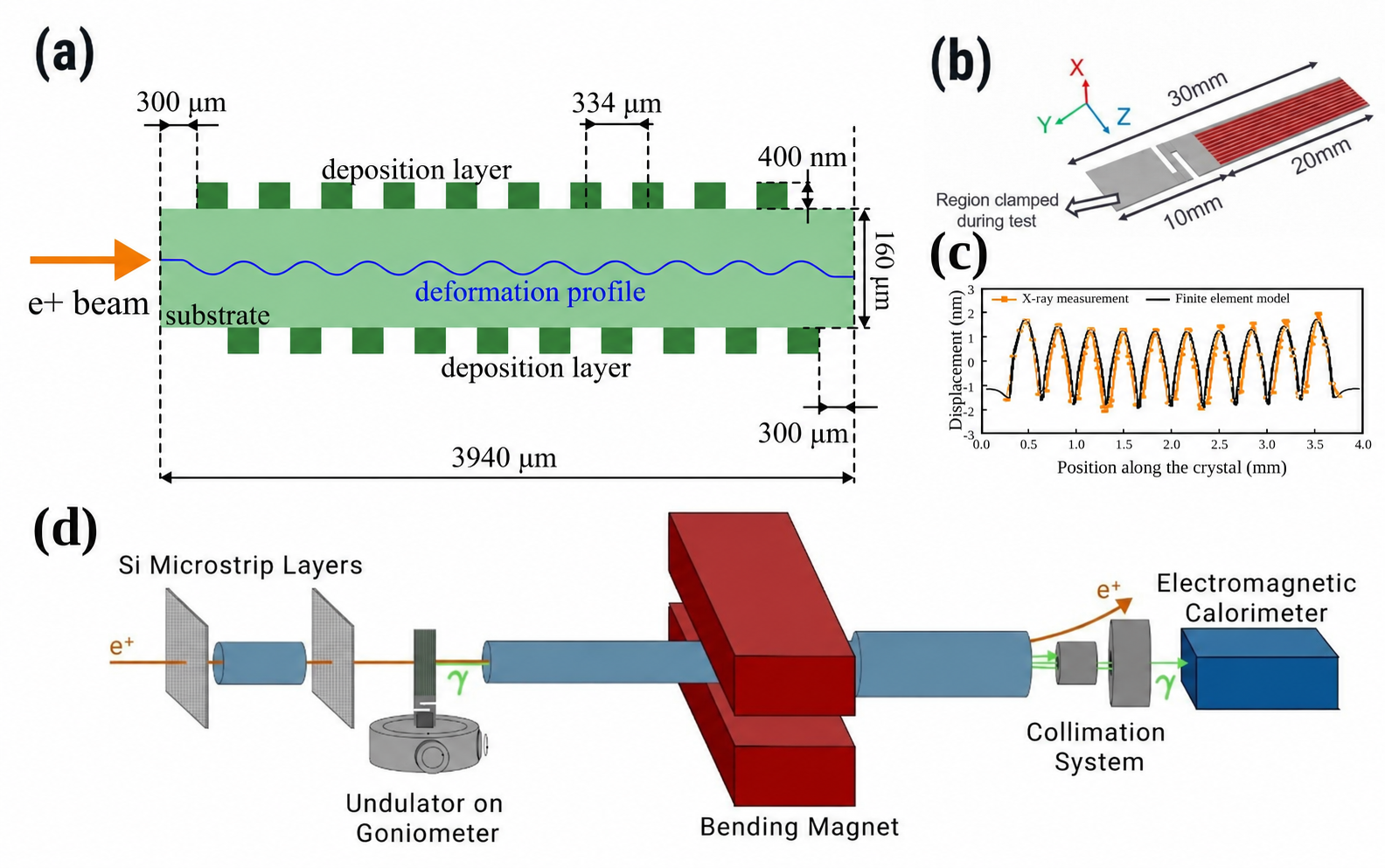}
\caption{(a) Cross section of the stressor-engineered CU. Alternating Si$_3$N$_4$ strips on opposite faces of the $160~\mu$m-thick Si(110) substrate generate ten periods with $\lambda_u=334~\mu$m. (b) Sample geometry. (c) Near-surface displacement profile measured by $8$~keV high-resolution X-ray diffraction and calculated by finite-element modeling. (d) Experimental layout at the CERN SPS H2 beam line. Distances not to scale.}
\label{fig:setup}
\end{figure*}

For a periodically bent trajectory, the energy of the $n$th harmonic observed at a polar angle $\vartheta$ is
\begin{equation}
E_{\gamma,n}(\vartheta)=
\frac{4\pi\hbar c\,\gamma^2 n}
{\lambda_u\left(1+\gamma^2\vartheta^2+K_{\rm eff}^2/2\right)}
\label{eq:undulator}
\end{equation}
with, for an approximately harmonic planar potential,
\begin{equation}
K_{\rm eff}^2\simeq
\left(\frac{2\pi\gamma A_u}{\lambda_u}\right)^2+
\frac{2\gamma U_0}{3m_ec^2}
\label{eq:keff}
\end{equation}
The first term represents the transverse motion imposed by the periodically bent planes, whereas the second accounts for the intrinsic channeling oscillations~\cite{Korol1999,Korol2014}. 

The CU was fabricated by patterning a $400$-nm tensile film of Si$_3$N$_4$ in strips on opposite faces of a $160~\mu$m-thick Si(110) wafer as originally proposed in Ref.~\cite{Guidi2007} and recently optimized in Ref.\cite{Malagutti2025}. The residual stress arises from differential thermal contraction after deposition. The realistic bending profile developed in \cite{Malagutti2025} was included in the simulations~\cite{Negrello2026} of the radiation spectra based on the channeling module ~\cite{sushko-2013} of the MBN Explorer software package ~\cite{solovyov2012}, showing the possibility to observe an undulator peak. 

The device employed in this study was fabricated using a refined version of the technique described in Refs. ~\cite{Guidi2007,Malagutti2025}. The detrimental effects of higher spatial harmonics in the bending profile were analysed in Ref ~\cite{Kostyuk2008}, where a sinusoidal surface deformation was proposed as a means of suppressing them. Here, we provide a practical fabrication solution based on a two-dimensional stressor pattern optimized to produce a nearly sinusoidal bending profile throughout the crystal thickness. This stressor-engineering method is the subject of a patent application~\cite{Patent2024}. The strip geometry was optimized by finite-element modeling (FEM) to produce an approximately sinusoidal deformation across the crystal thickness while suppressing higher spatial harmonics (see Fig.~\ref{fig:setup}(a,b)).

High-resolution X-ray diffraction at $8$~keV independently probed the periodic displacement. Figure~\ref{fig:setup}(c) compares the measured near-surface profile with the FEM prediction using the COMSOL code at the corresponding sampling depth. Their agreement supports the use of FEM to infer the deformation throughout the volume crossed by the positrons. At the crystal midplane, the calculated profile is nearly sinusoidal, with $A_u\simeq1.85$~nm and $\lambda_u=334~\mu$m. A weak long-wavelength curvature, attributed to a small imbalance between the stressor layers on the two surfaces, is retained in the deformation used for the radiation simulations.
For the measured deformation and a $35$~GeV positron beam, the bending contribution corresponds to an undulator parameter $K_u=2\pi\gamma A_u/\lambda_u\simeq2.38$, yielding $K_{\rm eff}\simeq2.77$ when the channeling contribution is included. Eqs.~(\ref{eq:undulator}) and (\ref{eq:keff}) predict an on-axis fundamental energy of
\begin{equation}
E_{\gamma,1}(0)\simeq7.2~\mathrm{MeV}
\label{eq:fundamental}
\end{equation}

These parameters place the device well within the large-amplitude, long-period regime.
The bending amplitude is nearly one order of magnitude larger than the Si(110)
interplanar spacing, $A_u/d_{110} \simeq 9.6$. The long-period condition can be
quantified from the Lindhard critical angle for planar channeling,
$\theta_c = \sqrt{2U_0/(pv)} \simeq \sqrt{2U_0/E}$, which for $35$~GeV positrons in
Si(110) and $U_0 \simeq 22$~eV gives $\theta_c \simeq 35~\mu$rad. In the harmonic
approximation, the corresponding channeling-oscillation period is
$\lambda_{\rm ch} \simeq \pi d_{110}/\theta_c \simeq 17~\mu$m, so that
$\lambda_u/\lambda_{\rm ch} \simeq 20$.

The experiment was performed at the CERN SPS H2 beam line [Fig.~\ref{fig:setup}(d)]. Two upstream double-sided silicon microstrip detectors~\cite{lietti-2013} reconstructed each incident positron trajectory, and the crystal was mounted on a precision goniometer. Downstream, a dipole magnet swept away charged particles, and the forward radiation passed through two coaxial 40-cm-long collimators (3~mm and 10~mm apertures) before reaching a 23-cm-long trapezoidal BGO calorimeter ($2.1$ and $2.9~\text{cm}^2$ front and back faces). The first collimator, $28.26$~m from the crystal, defined an angular half-acceptance of $53~\mu$rad~$\simeq 3.6/\gamma$. The planar channeling orientation was found with a 120~GeV electron beam, whose smaller critical angle allows a more precise alignment, by rotating the crystal while recording the full radiation spectrum.

A differential measurement was performed using two orientations of the same crystal. In the planar-channeling configuration, the periodically bent Si(110) planes were aligned with the incoming positron beam. In the reference configuration, the crystal was rotated by $1.6$~mrad, approximately $45\theta_c$. Identical incident phase-space selections were applied to the two data sets: reconstructed tracks were required to cross a common fiducial region and to enter within $\pm\theta_c$ of the mean beam direction.

\begin{figure}[!tbp]
\centering
\includegraphics[width=\columnwidth]{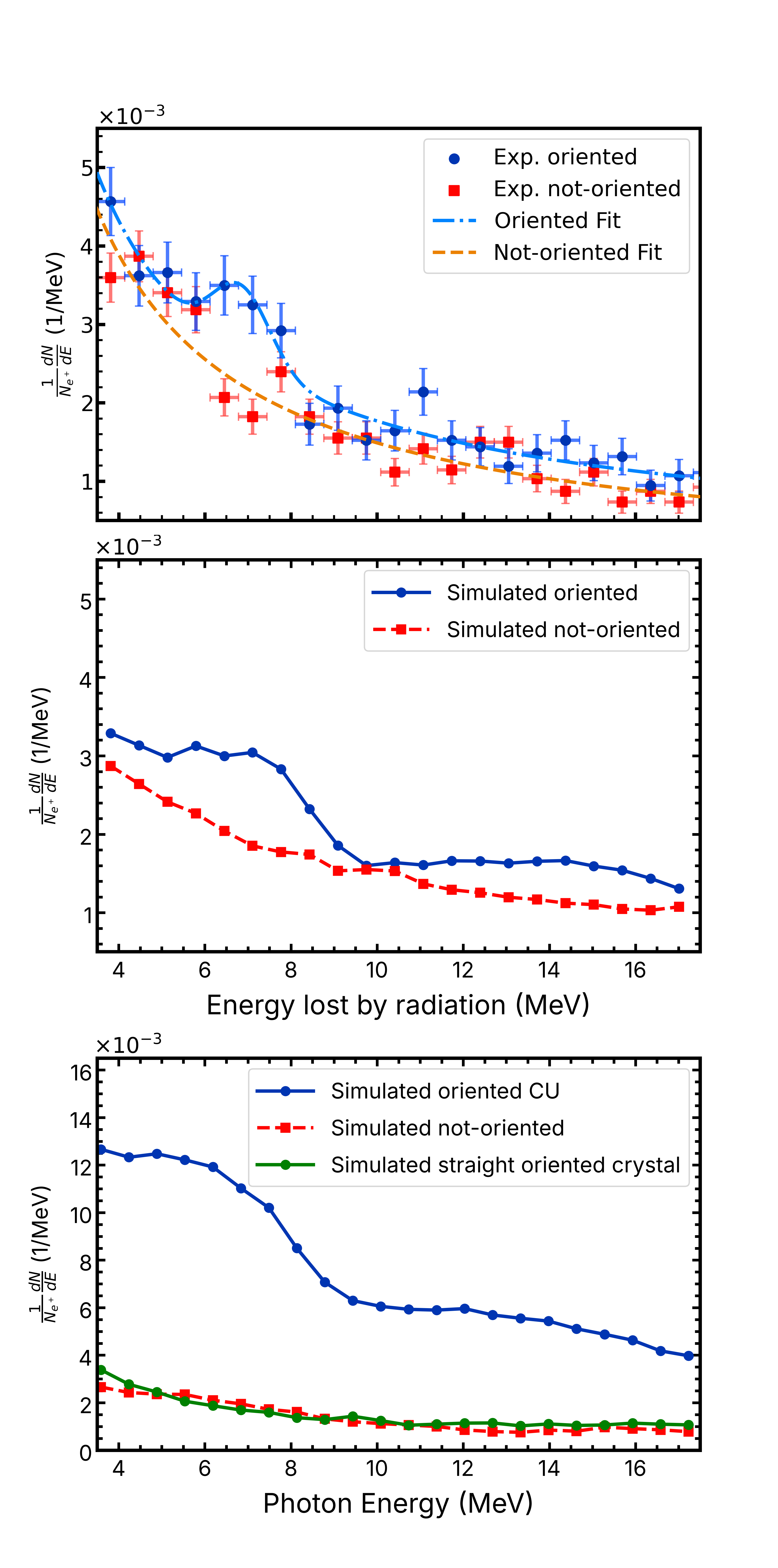}
\caption{Top: measured forward radiative-energy-loss distributions for planar alignment and the misaligned reference, normalized to the number of incident positrons. An excess around 7 MeV appears only in planar alignment. Center: simulated distributions for the FEM-derived CU deformation and for the misaligned crystal. Bottom: simulated single-photon spectra for the aligned and misaligned CU and for a straight crystal of the same thickness. The simulated collimation matches the experiment.}
\label{fig:spectra}
\end{figure}

Figure~\ref{fig:spectra} (top) shows the energy recorded by the forward BGO calorimeter for each positron, on an event-by-event basis. Since the detector integrates the energy carried by all photons entering its angular acceptance, the distributions represent total forward radiative energy loss rather than an inclusive single-photon spectrum. Owing to the restricted forward acceptance, however, spectral structures in the emitted radiation remain visible.
Both configurations exhibit a smoothly falling continuum. In planar alignment, a pronounced enhancement emerges between about $6$ and $8$~MeV, and the yield remains slightly higher up to the end of the displayed range. 

To test the CU interpretation, we performed detector-level Monte Carlo simulations with Geant4 \cite{Allison2016}, using the G4ChannelingFastSimModel \cite{Sytov2019,Sytov2023} for particle
transport and radiation in the oriented crystal and that recently was extended to the case of CU \cite{Sytov2023,Agapov2026}. Radiation emission is evaluated within the Baier--Katkov quasiclassical formalism
\cite{Guidi2012,RADCHARM2015}. The FEM-derived deformation profile of the fabricated crystal was implemented in the simulation. The incident positron phase space was sampled from the tracker data, including its resolution, while the experimental collimation and calibrated BGO response were included in
the detector-level simulation. The planar-channeling simulation reproduces the position and overall shape of the observed peak excess, whereas no localized feature remains in the misaligned configuration [Fig.~\ref{fig:spectra}, center]. Below approximately $5$~MeV, the measured yield exceeds the simulation in both configurations, possibly because of residual beam-line background not included in the model. 
The simulation also predicts a broader second-harmonic contribution at higher energies, for which only a weak indication is present in the data due to limited statistics.

Figure~\ref{fig:spectra} (bottom) shows the simulated single-photon spectra for both the oriented CU and the misaligned configuration. In the CU case the photon spectrum is strongly enhanced, whereas the enhancement is less pronounced in the energy-deposition distribution. This happens because the intense channeling radiation in the harder part of the spectrum shifts events towards higher total energy losses, which depletes the soft part of the energy-loss distribution. We then repeated the simulation for a straight Si crystal of the same thickness in the same planar-channeling configuration, keeping the beam, channeling, collimation and detector conditions identical. In this case the localized structure near $7$~MeV disappears, and the spectrum is close to that of the misaligned crystal. This indicates that the feature cannot be attributed to conventional channeling radiation alone, but requires a periodically bent crystal. Indeed, for the present parameters, planar channeling radiation features a broad spectrum peaking in the hundred-MeV range and contributes negligibly in the $4$--$17$~MeV region.

To quantify the excess, the deposited-energy spectrum measured in the misaligned configuration (Fig.~\ref{fig:spectra}, top) was fitted with a smooth continuum, 
$f_{\mathrm{mis}}(E) = a/E + b$. The aligned spectrum was fitted with the same continuum plus a Gaussian of mean $\mu$ and standard deviation $\sigma_{\mathrm{G}}$, providing a phenomenological
parametrization of the enhancement. The fit yields $\mu = 6.8 \pm 0.3$~MeV and
$\sigma_{\mathrm{G}} = 0.7 \pm 0.3$~MeV.
Under the null hypothesis that, in the absence of an undulator contribution, the aligned continuum coincides with the misaligned continuum, the aligned yield was compared bin by bin with $f_{\mathrm{mis}}$ within the window $[4.5,9]$~MeV. This interval was determined a priori, based on Geant4 prediction. 
The comparison used a $\chi^2$ test with six degrees of freedom, accounting for both Poisson counting uncertainties and uncertainties in the fitted background. The resulting p-value corresponds to a significance of $4.05\sigma$.

The measured peak position agrees with both the Geant4 simulation and the
$7.2$~MeV fundamental predicted from the XRD-validated deformation model, which
fixes $A_u$ and $\lambda_u$ independently of the radiation data. The slight
downward shift and broadening of the peak arise from the finite resolution, the
wide collimation angle, and the calorimetric measurement of the total forward
energy, all included in the simulation.

\begin{figure}[t]
\centering
\includegraphics[width=\columnwidth]{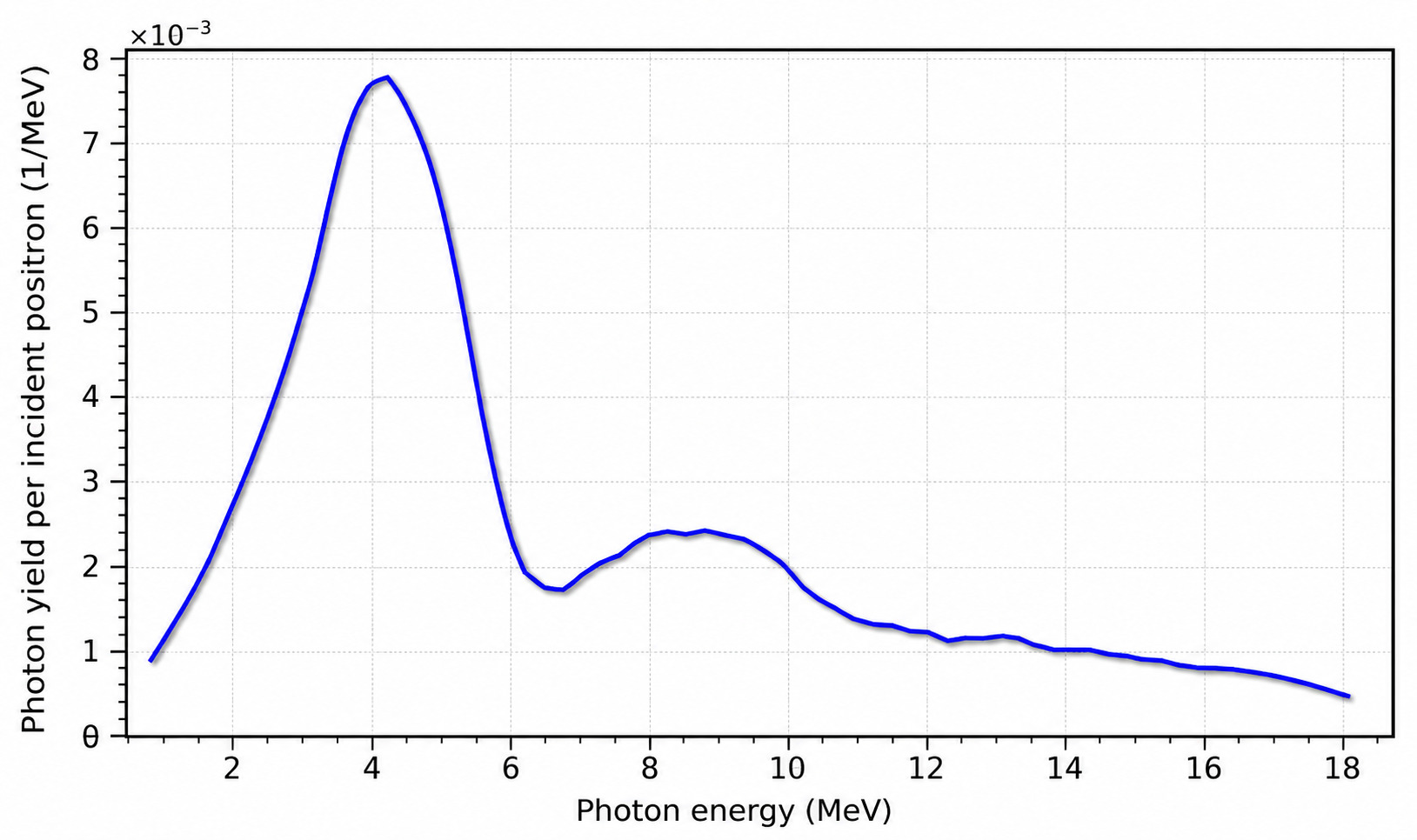}
\caption{Simulated photon yield per incident $20$~GeV positron from the fabricated ten-period CU for a photon-collection half-angle $\theta_{\rm col}=1/\gamma \simeq 25~\mu$rad. The reduced angular acceptance reveals a pronounced fundamental above $4$~MeV.}
\label{fig:fcc}
\end{figure}

The $53~\mu$rad acceptance, approximately $3.6/\gamma$, integrates photons emitted at different angles and hence, according to Eq.~(\ref{eq:undulator}), at different energies, which broadens the observed fundamental. To illustrate the spectrum for a collection geometry typical of undulator sources, we simulated the same ten-period crystal for the $20$~GeV positron beam foreseen at the FCC-ee injector~\cite{Agapov2026}, with an angular half-acceptance of $1/\gamma \simeq 25~\mu$rad. As shown in Fig.~\ref{fig:fcc}, the reduced acceptance yields a pronounced fundamental above $4$~MeV together with a weaker higher-harmonic structure. Using the nominal beam intensity and transverse phase-space parameters of Ref.~\cite{Agapov2026}, the simulated peak spectral brilliance would be approximately $1.8\times10^{21}$~photons~s$^{-1}$~mm$^{-2}$~mrad$^{-2}$~(0.1\%~BW)$^{-1}$, comparable to the design peak brilliance of advanced inverse-Compton gamma-ray sources in the MeV domain~\cite{Krafft2010,Petrillo2023,Adriani2014,Piersanti2018} and exceeding that of operating facilities such as HI$\gamma$S~\cite{Weller2009}.

The $20$~GeV projection illustrates how the experimentally tested stressor-engineering concept can be extended beyond the parameters of the present device. The $\lambda_u=334~\mu$m period realized here should be regarded as an experimentally tested operating point rather than a technological limit. The lithographic stressor approach provides a straightforward route to extending the present ten-period pattern to several tens of periods, potentially yielding a narrower and more intense harmonic structure, with the useful number of periods primarily constrained by dechanneling. Shorter bending periods provide a complementary route: when $A_u$ is co-optimized to control $K_{\rm eff}$, they can shift the fundamental toward higher photon energies or allow a given photon energy to be reached at lower beam energy. They can also increase the number of oscillations within the available channeling length, provided that the increased curvature does not substantially shorten the dechanneling length. For positrons this constraint is much less severe than for electrons, which makes the positron-based LALP scheme a natural route toward thicker crystals with more undulator periods and higher brilliance. Achieving the large-amplitude regime at shorter periods requires stronger bending stresses, for which coactive thin films provide a promising route~\cite{Hamza2026}. Because the active bulk itself is periodically deformed, no separate straight backing section, such as that used in the recent diamond-superlattice realization~\cite{Backe2026}, is required, avoiding a distinct channeling-radiation background.

In summary, our measurement provides evidence for CU radiation
from $35$~GeV positrons channeled through a ten-period, stressor-engineered Si(110)
crystal, quantitatively consistent with the engineered crystal geometry. The planar-channeling distribution exhibits a localized excess at
$6.8\pm0.3$~MeV, in agreement with Monte Carlo simulations and consistent with the
$7.2$~MeV fundamental predicted analytically, with a statistical significance of
$4.05\sigma$. No corresponding feature is observed after crystal misalignment, and in
simulations it disappears when the periodic bending is removed while planar channeling
is retained. Owing to the longer dechanneling length of positrons, this approach opens a favorable route toward thicker CUs with more periods and, as projected for the FCC-ee injector, a peak brilliance comparable to inverse-Compton sources. These results establish a quantitative link between engineered crystal geometry and tunable MeV radiation.

\begin{acknowledgments}
This work was supported by the EU Commission through the EIC Pathfinder Open project TECHNO-CLS (GA No. 101046458) and by the INFN CSN5 projects CORAL and GEANT4INFN. We acknowledge ISCRA for access to the LEONARDO supercomputer, owned by the EuroHPC Joint Undertaking, hosted by CINECA (Italy). We gratefully acknowledge the CERN SPS and PS accelerator and beam-line staff and the CERN Geodetic Metrology (BE-GM) team for their excellent support during the experiment. The stressor-engineering technology used in this work is the subject of Italian patent application IT~102024000022335, jointly owned by INFN, the University of Padua, and the University of Ferrara~\cite{Patent2024}. Generative AI tools were used solely to improve the English language and readability of the manuscript. The authors take full responsibility for its scientific content. 
\end{acknowledgments}

\bibliography{references2}

\end{document}